\documentstyle{article}
\begin{document}
\centerline{\Large \bf Reply to} \vskip 0.1in \centerline{\Large
\bf No Contradictions between Bohmian} \vskip 0.1in
\centerline{\Large \bf	and quantum mechanics} \vskip 0.2in
\centerline{Partha Ghose} \vskip 0.1in \centerline{\it S. N. Bose
National Centre for Basic Sciences, III/JD, Salt Lake, Calcutta
700 098, India} \vskip 0.4in

In a recent paper\cite{Marchildon} Marchildon has claimed that the
basic conclusion of my papers\cite{pg1,pg2} regarding the
incompatibility of Bohmian mechanics and standard quantum
mechanics is unfounded. This is based on a simple mistake. Let us
consider Marchildon's final equation (18) which he uses to refute
my claim. Let $y = x_1 + x_2$. Then eqn.(18) can be written as
\begin{equation}
\frac{d y}{d t} = c y,
\end{equation}
where $c = \frac{\hbar k}{m L}$. Integrating this we get
\begin{equation}
y(t) = y(0) e^{c t}.
\end{equation}
Since $y(0) = x_1(0) + x_2(0)= 0$ by assumption, it follows that
$y(t)=0$ for all $t$. Q. E. D.

As would be clear from my papers, the conditions under which
Fraunhoffer diffraction holds was essential for the conclusion I
drew. If one carries out these approximations in Marchildon's
calculations, the quadratic terms in $a$ and $x$ must be
dropped\cite{bw} in his equations (13) and (14) which therefore
reduce to
\begin{eqnarray}
r_A &\approx& y - \frac{2 a x}{2 L}\\ r_B &\approx& y + \frac{2 a
x}{2 L}
\end{eqnarray}
When these are substituted into his equation (17), the right hand
side of his equation (18) vanishes.

What Marchildon's calculations have, in fact, established is that
the assumption of translation invariance (Fraunhoffer diffraction
and plane waves) imposed by me in my papers are sufficient but not
necessary to obtain the basic result, namely that the trajectories
are symmetrical at all times and do not cross, and therefore imply
the incompatibility of Bohmian mechanics and standard quantum
mechanics, as I claimed.

\end{document}